\documentstyle[aps,epsfig,prd,preprint]{revtex}
\tightenlines

\def \beq{\begin{equation}}         \def \eeq{\end{equation}}
\def \beqa{\begin{eqnarray}}        \def \eeqa{\end{eqnarray}}
\def \bea{\begin{array}}        \def \eea{\end{array}}

\def\nomb{\nonumber}
\def \abs#1{\left| #1 \right|}
\def \lb{\left(}  \def \rb{\right)}
\def\npb#1#2#3{    {\it Nucl. Phys. }{\bf B#1} (#2) #3}

\def\plb#1#2#3{    {\it Phys. Lett. }{\bf B#1} (#2) #3}
\def\prd#1#2#3{    {\it Phys. Rev. }{\bf D#1} (#2) #3}

\def\prl#1#2#3{    {\it Phys. Rev. Lett. }{\bf #1} (#2) #3}

\def\epj#1#2#3{{\it Eur. Phys. J.}{\bf C#1}(#2)#3}

\begin{document}
\draft
\title{ Implication of $\sin 2\beta$ from global fit and $B\to J/\psi K_S$ }
\author{{ Yue-Liang Wu and Yu-Feng Zhou}\\
{\small \ Institute of Theoretical Physics, Chinese Academy of
Science, Beijing 100080, China }}
\maketitle

\begin{abstract}
The measurement of $\sin 2 \beta$ is discussed within and beyond
the standard model.  In the presence of new physics, the angle
$\beta$ extracted from the global fit(denoted by
$\beta^{SM}_{fit}$) and the one extracted from $B\to J/\psi K_S$
(denoted by $\beta_{J/\psi}$) are in general all different from
the 'true' angle $\beta$ which is the weak phase of CKM matrix
element $V^*_{td}$. Possible new physics effects on the ratio
$R_{\beta}=\sin2\beta_{J/\psi}/\sin 2\beta^{SM}_{fit}$ is studied
and parameterized in a most general form. It is shown that the
ratio $R_{\beta}$ may provide a useful tool in probing new
physics.  The experimental value of $R_{\beta}$ is obtained
through an update of the global fit of the unitarity triangle with
the latest data and found to be less than unity at 1$\sigma$
level.  The new physics effects on $R_{\beta}$ from the models
with minimum flavor violation (MFV) and the standard model with
two-Higgs-doublet (S2HDM) are studied in detail.  It is found that
the MFV models seem to give a relative large value $R_{\beta}\geq
1$. With the current data, this may indicate that this kind of new
physics may be disfavored and alternative new physics with
additional phases appears more relevant. As an illustration for
models with additional phase beyond CKM phase, the S2HDM effects
on $R_{\beta}$ are studied and found to be easily coincide with
the data due to the flavor changing neutral Higgs interaction.
\end{abstract}

\newpage

\section{Introduction}
Recently, the BaBar and Belle Collaborations have reported their new
results on the measurements of   time dependent CP
asymmetry ${\cal A}_{CP}$ of decay mode $B\to J/\psi K_S$. The
definition of ${\cal A}_{CP}$ is given as follows
\beqa
{\cal A}_{CP}(t) &=&  { \Gamma(\bar{B}^0(t)\to J/\psi K_S) - \Gamma(B^0(t)\to J/\psi K_S)  \over
                \Gamma(\bar{B}^0(t)\to J/\psi K_S) +\Gamma(B^0(t)\to J/\psi K_S)}\nomb \\
&\equiv& -\sin 2\beta_{J/\psi} \sin ( \Delta m_B t).
\eeqa

In the framework of the Standard Model (SM), the angle
$\beta_{J/\psi}$ is expected to be equal to the angle $\beta$
which concerns the weak phase of Cabbibo-Kabayashi-Maskawa (CKM)
matrix element $V^*_{td}$ \beqa \beta \equiv -arg \left( { V_{td}
V^*_{tb}  \over V_{cd} V^*_{cb} }\right) \eeqa
 which is also one of the angles of the unitarity triangles
containing $d$ and $b$ quarks. The new measurements give the
following values \cite{babar,belle}:
 \beqa
 \sin 2\beta_{J/\psi}|_{exp} &=& 0.59\pm 0.14\pm 0.05 \mbox{\ \ (BaBar)}\nomb \\
 \sin 2\beta_{J/\psi}|_{exp} &=& 0.58^{+0.32+0.09}_{-0.34-0.1} \mbox{\ \  (Belle)}.
 \eeqa
 Combining with the ones measured earlier by CDF and ALEPH Collaboration
 $0.79^{+0.41}_{-0.44}$\cite{CDF}, and $0.84^{+0.82}_{-1.04}\pm 0.16 $\cite{ALEPH},
 one arrives at the new world average value of $\sin 2\beta$

\beqa \sin 2\beta_{J/\psi}|_{exp} = 0.59\pm 0.14 . \eeqa

Although this result is consistent with the prediction  of
Standard Model (SM),
the results from BaBar and Belle may imply a
possibility that the angle $\beta_{J/\psi}$ measured directly from the
time dependent CP asymmetry may be different than the  one from the global
fit in SM ( denoted by $\beta^{SM}_{fit}$) which contains the indirect
CP violations of both neutral $K$ and $B$ meson mixings.  This
possibility has aroused many discussions in the
literatures\cite{other,other1,other2,nir,buras1,buras2,buras3,Ali}. As has been
pointed out in Ref.\cite{nir}, a lower value of $\sin 2\beta$ less
than 0.5 may imply that some hadronic parameters are out of the
reasonable range. It may imply a large bag parameter $\hat{B}_K$ for
the matrix element $<K^0| H_{W}|\bar{K}^0>$, a large $SU(3)$ breaking
factor between $B^0$ and $B^0_s$ or a small value of $|V_{ub}|$.
In the SM, a lower bound of $\sin 2\beta$ can be derived from the
evaluation of  $\epsilon_K$ and $\Delta m_B$, a conservative bound
was found to be $\sin 2\beta>0.42$ \cite{buras1} which is compared
with $0.34$ in\cite{buras2} due to the use of different values of
hadronic parameters. The existence of a lower bound on $\sin
2\beta$ also holds for new physics models with minimal flavor
violation (MFV), namely a class of models which has no new flavor
changing operators beyond those in the SM and no additional weak
phases beyond the CKM phase.

In the SM, the angle $\beta^{SM}_{fit}$ extracted from the global
fit should be the same as the one measured from the time dependent
CP asymmetry in the decay $B\to J/\psi K_S$.  However, if there
exists new physics beyond the SM, the situation may be quite
different, the angle ``$\beta$'' extracted from two different
approaches are in general not equal. This is because $\epsilon_K$
and $\Delta m_B$ as well as $B\to J/\psi K_S$  will receive
contributions from new physics in a quite different manner
comparing to the ones from the SM. As a consequence, the extracted
``$\beta$'' from two ways will become different. In the most
general case, one may find that $\beta^{SM}_{fit} \neq
\beta_{J/\psi}\neq \beta$ when new physics exists and the ratio
\beqa
R_{\beta}\equiv{\sin 2\beta_{J/\psi} \over \sin
  2\beta^{SM}_{fit}}
\eeqa
is therefore not equal to unity. It will be
shown in detail below that the true angle $\beta$ may not be
directly obtained from the measurement of $\beta^{SM}_{fit}$ or
$\beta_{J/\psi}$.

 It is obvious that the deviation from $R_{\beta}=1$ provides a clean signal
of new physics. While the different new physics models may modify
its value in different ways.  In the definition of $R_{\beta}$,
the value of $\sin 2\beta_{J/\psi} $ is model dependent, the
prediction of $\sin 2\beta_{J/\psi} $ may vary largely from
different new physics models. On the other hand, even when
$\sin2\beta^{SM}_{fit}$ is extracted via the same formulae as the
ones in the SM and its value only depends on the hadronic
parameters and the experimental data, namely it looks like model
independent, but possible existence of new physics implies that
$\beta^{SM}_{fit}$ may not be necessarily equal to the true
$\beta$ in the SM. This is because the experimental data should
contain all contributions from both SM and beyond the SM if new
physics truly exists. Thus the relation between $\beta^{SM}_{fit}$
and $\beta$ depends on models. In this paper, we present a general
parameterization for possible new physics contributions to
$R_{\beta}$. For a detailed consideration, we investigate two
interesting and typical models, one is the model with MFV and
another is the simplest extension of SM with just adding one Higgs
doublet. For convenience of mention in our following discussions,
we may call such a minimal extension of the standard model with
two Higgs doublet as an
S2HDM\cite{TH1,TH2,TH3,TH4,WW,YLW,Wu-Zhou}, in which CP violation
could solely originate from the Higgs potential\cite{TDL,WW,YLW}.
The experimental value of $R_{\beta}$ is obtained from an update
of the global fit for the unitarity triangle (UT). It will be seen
that the ratio $R_{\beta}$ may provide a useful tool for probing
new physics.  The current data lead to a low value of $R_{\beta}$
which is not likely to be accommodated by the models with MFV. If
the future experiments confirm such a low value of $R_{\beta}$,
the models involving new interactions with additional phases
beyond the CKM phase will be preferred. Taking the S2HDM as an
example, we illustrate that the new physics models with additional
CP phases can easily explain the current data.

This paper is organized as follows: in section {\bf I$\!$I}, the ratio
$R_{\beta}$ with possible new physics effects is introduced and the new
physics effects on $R_{\beta}$
are parameterized in a most general form.  In section {\bf
I$\!$I$\!$I}, the profile of the unitarity triangle is updated with the latest data
and the resulting value of $R_{\beta}$ is obtained. The new physics
effects of the models with MFV and the S2HDM are discussed in section {\bf I$\!$V}.  Our
conclusions are made  in the last section.

\section{ Neutral meson mixing within and beyond the SM}
\begin{figure}
\centerline{ \psfig{figure=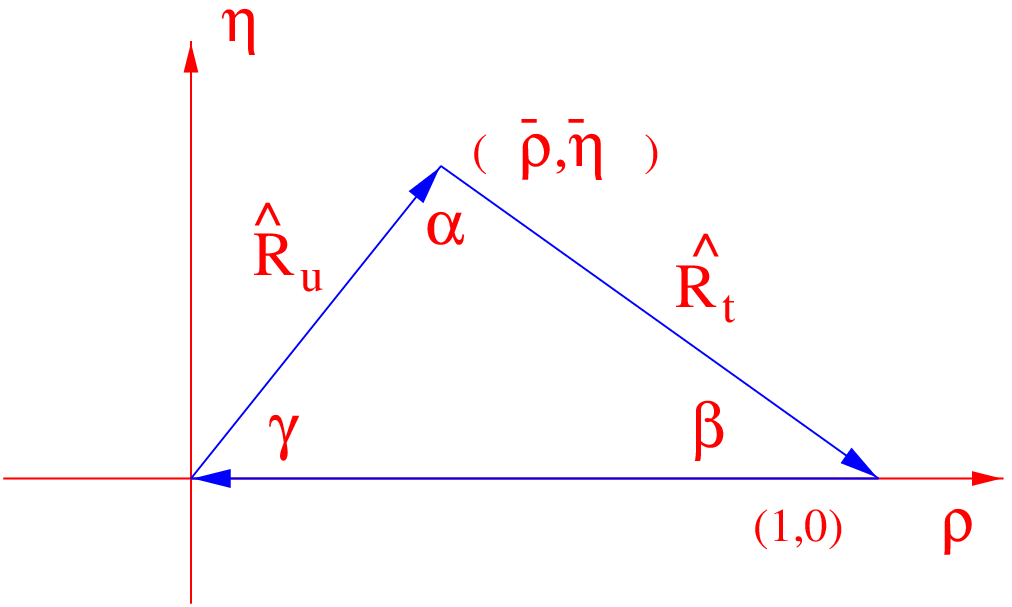, width=12cm} }
 \caption{The unitarity triangle}
\label{triangle.eps}
\end{figure}

In the study of quark mixing and CP violation, it is convenient to
use the Wolfenstein parameterization \cite{wolfen} in which the
CKM matrix elements can be parameterized by four parameters
$\lambda$, $A$, $\rho$ and $\eta$. The unitarity relation \beqa
V_{ud}V^*_{ub}+V_{cd}V^*_{cb}+V_{td}V^*_{tb}=0 \label{unitarity}
\eeqa can be depicted as a closed triangle in the  $\rho$, $\eta$
plane (see Fig.\ref{triangle.eps}). The  two sides of the triangle
can be written as:

\beqa
\hat{R}_{u} &\equiv& -\frac{V_{ud}V^*_{ub}}{V_{cd}V^*_{cb}}
     =R_u e^{i\gamma}
     =\sqrt{\bar{\rho}^2+\bar{\eta}^2}  e^{i\gamma}, \nomb \\
\hat{R}_{t} &\equiv& -\frac{V_{td}V^*_{tb}}{V_{cd}V^*_{cb}}
     =R_t e^{-i\beta}
     =\sqrt{(1-\bar{\rho})^2+\bar{\eta}^2}  e^{-i\beta},
\eeqa
with $\bar{\rho}\equiv (1-\frac{\lambda^2}{2})\rho$, and
$\bar{\eta}\equiv (1-\frac{\lambda^2}{2})\eta $.
The three angles of  $\alpha$, $\beta$ and $\gamma$ of the triangle are defined as follows
\beqa
\sin 2\alpha &=& {      2 \bar{\eta} (  \bar{\eta}^2 +\bar{\rho}^2-\bar{\rho})
                    \over (\bar{\rho}^2+ \bar{\eta}^2)((1-\bar{\rho})^2+ \bar{\eta}^2)}, \nomb\\
\sin 2\beta  &=& {2 \bar{\eta}(1-\bar{\rho}) \over
(1-\bar{\rho})^2+ \bar{\eta}^2 }, \nomb \\ \sin 2\gamma &=& { 2
\bar{\rho}\bar{\eta} \over \bar{\rho}^2+ \bar{\eta}^2 },
\label{angle-def}
\eeqa
with $\alpha+\beta+\gamma=180^\circ$. One
of the important goals in the modern particle physics is to
precisely determine those angles and to test the unitarity
relation of Eq.(\ref{unitarity}). Many efforts have been done for
that purpose and the current data have constrained the apex of the
triangle in a small region.  For example, the values of $\lambda$,
$|V_{cb}|$ and $|V_{ub}|$ are extracted from the semileptonic $K$
and $B$ decays. Besides the semileptonic decays, the important
constraint may come from the neutral meson mixings, such as
$K^0-\bar{K}^0$ and $B^0_{(s)}-\bar{B}^0_{(s)}$ mixings.

There are  two important observables in neutral meson mixings:
the indirect CP violating parameter $\epsilon_K$ of kaon and
the mass difference of neutral $B$ meson  $\Delta m_{B_{(s)}} $.
In the SM, their expressions read
\beqa
\epsilon^{SM}_K&=&\left\{\bar{\eta}
\left[(1-\bar{\rho})A^2 \eta_2 S_0(x_t)+P_{\epsilon}\right]
A^2 \hat{B}_K C_{\epsilon}\right\}e^{i {\pi \over 4}}
\label{SMepsilon} \nomb\\
\Delta m^{SM}_{B_{(s)}}&=&{G^2_F \over 6 \pi^2 } m^2_W m_{B_{(s)}}
\left(f_{B_{(s)}}\sqrt{B_{(s)}}\right)^2
\eta_B S_0(x_t) |\lambda^{B_{(s)}}_t|^2 .
\eeqa
where $S_0(x_i)$ with $(x_i=m^2_i/M^2_W)$ are the integral
function arising from the box diagram\cite{I-L}  and $\eta_i$s are the QCD
corrections with the values \cite{S0}
$\eta_1=1.38\pm0.20, \; \; \eta_2=0.57\pm0.01,  \; \; \eta_3=0.47\pm0.04$
and $ \eta_B=0.55\pm 0.01$,
$G_F=1.16\times 10^{-5} $ GeV$^{-2}$ is the Fermi coupling constant, $f_B$ and $B_K, B$ are
the decay constant and bag parameter for kaon and $B$ meson respectively.
The two constants
$C_\epsilon$ and $P_\epsilon$ have the values
\cite{buras1,buras2}:
\[
C_{\epsilon}={G_F^2 F^2_K m_K m^2_W }\lambda^{10}/({6\sqrt{2}\pi^2 \Delta m_K})
            =0.01
\] and
\[
P_{\epsilon}=\left[\eta_3 S_0(x_c,x_t)-\eta_1 S_0(x_c)\right]/{\lambda^4}
            =0.31\pm 0.05
\]

In the global SM fit of UT, one usually assumes that the SM is
fully responsible for the observed experimental data.  This
scenario, however may be invalid if there are new contributions
beyond the SM. If new physics effects exist, then
$\epsilon_K^{exp.} \neq \epsilon_K^{SM}(\bar{\rho},\bar{\eta})$
and
 $\Delta m^{exp}_{B}\neq \Delta m^{SM}_{B}(\bar{\rho},\bar{\eta})$ . In
 general, one may write
 \beqa
 \epsilon_K^{exp.} & = &  \epsilon_K^{SM}(\bar{\rho},\bar{\eta}) [1 + \delta
 \epsilon_K ] \equiv
 \epsilon_K^{SM}(\bar{\rho}_{fit}^{SM},\bar{\eta}_{fit}^{SM}) \nonumber \\
 \Delta m^{exp}_{B} & = &  \Delta m^{SM}_{B}(\bar{\rho},\bar{\eta}) [1 + \delta m_B]
 \equiv \Delta m^{SM}_{B}(\bar{\rho}_{fit}^{SM},
 \bar{\eta}_{fit}^{SM})[1 + \delta \tilde{m}_B ]
 \eeqa
 where $ \delta\epsilon_K$ and $\delta m_B$ ( $\delta \tilde{m}_B$) represent possible new
 physics contributions. Therefore the parameters $\rho$ and $\eta$ extracted
from the fit are in general the effective ones which may
significantly deviate from their true values in the SM .  The
effective values $\rho^{SM}_{fit}$ and $\eta^{SM}_{fit}$ are
defined as follows:
 \beqa
& & \epsilon^{exp}_K = \left\{\bar{\eta}^{SM}_{fit}
\left[(1-\bar{\rho}^{SM}_{fit})A^2 \eta_2
S_0(x_t)+P_{\epsilon}\right] A^2 \hat{B}_K
C_{\epsilon}\right\}e^{i {\pi \over 4}} \label{SMepsilonfit}
 \nomb\\
 & &  \Delta m^{exp}_{B}/[1 + \delta \tilde{m}_B ] = {G^2_F \over  6 \pi^2 }
(1-{\lambda \over 2})^2 A^2 \lambda^4 m^2_W m_{B} \left( f_B
\sqrt{B}\right)^2 \eta_B S_0(x_t)
 \left( (1-\bar{\rho}^{SM}_{fit})^2+(\bar{\eta}^{SM}_{fit})^2 \right)  .
\eeqa
where, $\epsilon^{exp}_K$ and $\Delta m^{exp}_B$ are the values
measured from the experiments \cite{PDG}. It is seen that in the
existence of unknown new physics, one may even not be able to extract
the effective parameters $\bar{\rho}_{fit}^{SM}$ and
$\bar{\eta}_{fit}^{SM}$ from the data $\epsilon^{exp}_K$ and $\Delta
m^{exp}_{B}$.  Only when the new physics contributions to the term $
\delta \tilde{m}_B$ are neglected, the two effective parameters
$\bar{\rho}_{fit}^{SM}$ and $\bar{\eta}_{fit}^{SM}$ can be determined
from the data $\epsilon^{exp}_K$ and $\Delta m^{exp}_{B}$. In our
following fit,  due to the large uncertainties in both
theoretical parameters and the present data, we actually take $ \delta
\tilde{m}_B=0$ as an approximation, so that
the angle $\beta^{SM}_{fit}$ can be extracted
via the same form as the one in Eq.(\ref{angle-def})
 \beqa
  \sin 2\beta^{SM}_{fit}  \equiv {2
\bar{\eta}^{SM}_{fit}(1-\bar{\rho}^{SM}_{fit}) \over
(1-\bar{\rho}^{SM}_{fit})^2+ (\bar{\eta}^{SM}_{fit})^2 },
\label{SMfit-epsK-dmb}
 \eeqa
 From Eq.(\ref{SMfit-epsK-dmb}) , by eliminating $S_0(x_t)$ in the expression for $\epsilon_K$ and
$\Delta m_B$,  the following relation holds
 \beqa
 \sin 2\beta^{SM}_{fit}&=& C_K \left(1-\omega
\bar{\eta}^{SM}_{fit}\right). \label{beta-sm}
\eeqa
with
$C_K=\frac{G_F^2 m_B m_W^2 F_B^2 B_B \eta_B |\epsilon_K|}
               {6\pi^2\Delta m_B A^4 \hat{B}_K \eta_2 C_\epsilon}$   and
 $\omega = {C_\epsilon A^2 \hat{B}_K P_\epsilon /|\epsilon_K|}
$.

 Before proceeding, we would like to emphasize that even if taking the
 expressions satisfying the same relations as the ones in the SM, it is clear that
 the resulting values for $\rho^{SM}_{fit}$ and $\eta^{SM}_{fit}$ are not
 necessary to be the true ones in the SM if new physics beyond the SM truly exists.
 In general, one has $\rho^{SM}_{fit}\neq \rho$ and
$\eta^{SM}_{fit}\neq \eta$. Only when there are no new
 physics contributions to the $K^0-\bar{K}^0$ and
  $B^0-\bar{B}^0$ mixings, i.e. $\epsilon^{exp}_K=\epsilon^{SM}_K$
and $\Delta m_B^{exp}=\Delta m_B^{SM}$, one can then get the true
values $\bar{\rho}^{SM}_{fit}=\bar{\rho}$ and
$\bar{\eta}^{SM}_{fit}=\bar{\eta}$. Thus the fit can really probe
the true value of $\sin 2\beta$. However, if there exists new
 physics beyond the SM,  the relation between $\sin 2\beta^{SM}_{fit} $
 and $\sin 2\beta$ will become complicated and depend strongly on new physics
 models, which can explicitly be seen below. Note that
 the values of $\rho^{SM}_{fit}$ and $\eta^{SM}_{fit}$ fitted in
 such a way remain model independent though they may not be equal to
 the true values of $\rho$ and $\eta$ in the SM.

 It is noticed that in the SM neutral meson mixings only occur at loop level,
 new physics contributions are expected to be considerable.  If we
 denote the transition matrix element $<f^0|H_{eff}|\bar{f}^0>$ of
 neutral meson $f^0 (f^0=K^0,B^0_{(s)})$ by $M^f_{12}$, then the new
 physics effects on $M^f_{12}$ can be parameterized in the following
 general form:

\beqa
 M^f_{12}=\sum_{i,j=c,t} \lambda^f_i \lambda^f_j
M^{SM,f}_{12,ij}(1+r^f_{ij})
         + \tilde{M}^f_{12}
\label{def}
\eeqa
where $\lambda^f_i  ( i=c,t )$ are the products of CKM
matrix elements with the definitions
$
\lambda^{K}_i\equiv V^*_{id}V_{is} , \;
\lambda^{B}_i\equiv V^*_{id}V_{ib} , \; $
and $\lambda^{B_s}_i\equiv V^*_{is}V_{ib} .$
The matrix elements $M^{SM,f}_{12,ij}$ are the ones in SM. They
receive contributions from the internal $tt,cc$ and $ct$ quark
loops in the box diagrams which are responsible for the
$f^0-\bar{f}^0$ mixing.  Real parameters $r^f_{ij}$ reflect the
possible new physics contributions to the corresponding loop
diagrams, which carry no additional weak phases besides
$\lambda^f_{ij}$. Models with such property include some versions
of SUSY models and type I and I$\!$I 2HDM. Obviously those models
belong to the ones with MFV since they carry no additional phases.
New physics contributions with new weak phases are all absorbed
into the last term $\tilde{M}^f_{12}$. This term may arise from
additional new interactions, such as the tree level Flavor
Changing Neutral Current (FCNC) and the superweak-type
interactions\cite{LW1} as well as interacting terms with new weak
phases.

In evaluating $\epsilon_K$ in kaon system, although the
contribution of $tt$ quark loop is dominant, the effects of $cc$
and $ct$ quark loops are not negligible. While in the calculations
of the mass differences of $B^0_{(s)}$ meson, one may only
consider the $tt$ quark loop in the box diagram.

By using the Wolfenstein parameterization for CKM matrix elements
and the parameterization of Eq.(\ref{def}), in the presence of new
physics, the expression of
 $\epsilon_K$ is
\beqa
\epsilon_K=\left\{\bar{\eta}
\left[(1-\bar{\rho})A^2 \eta_2 S_0(x_t)\left(1+r^K_{tt}\right)+P_{\epsilon}(1+r_{ct})\right]
A^2 \hat{B}_K C_{\epsilon}\right\}e^{i {\pi \over 4}}
 +\tilde{\epsilon}_K
\label{newepsilon}
\eeqa
with
\beqa
r_{ct}&\equiv&{ \eta_3 S_0(x_c,x_t)r^K_{ct}- \eta_1 S_0(x_c)r^K_{cc}
         \over
         \eta_3 S_0(x_c,x_t)- \eta_1 S_0(x_c)}\nomb \\
\tilde{\epsilon}_K &\equiv& { e^{i{\pi\over4}} \over \sqrt{2}
\Delta m_K} \mbox{Im}\tilde{M}^K_{12}
 \eeqa
 where the new physics effects with MFV have been expressed as the
corrections to the integral function $S_0(x_t)$ and $P_\epsilon$.
The effects  with additional phases have been absorbed into
$\tilde{\epsilon}_K$.
The $B^0_{(s)}-\bar{B}^0_{(s)}$ mass difference is also modified as follows
\beqa
\Delta m_{B}&\simeq&\left| M^B_{12} \right|
=\Delta m^{SM}_{B}\left|1+r^{B}_{tt}\right|
    \abs{  e^{-2i\beta}+\tilde{r}^{B} e^{2i\phi^{B}}}.
\label{newdmb} \\
\Delta m_{B_s}&\simeq&\left| M^{B_s}_{12} \right|
=\Delta m^{SM}_{B_s}\left|1+r^{B_s}_{tt}\right|
    \abs{1+\tilde{r}^{B_s} e^{2i\phi^{B_s}}}.
\label{newdmbs}
\eeqa
 Where we have kept only the $tt$ quark loop in the box diagrams, the
contributions from other loops with internal $cc$ and $ct$ quarks are
highly suppressed and can be safely neglected.

 From the definition of ${\cal A}_{CP}$, the $\beta_{J/\psi}$
measured from the time dependent CP asymmetry will be the total
phase of $M^B_{12}$. Thus in the presence of new physics, The
angle $\beta_{J/\psi}$ extracted from the time dependent CP
asymmetry ${\cal A}_{CP}(t)$ in decay $B\to J/\psi K_S$ is in
general different from the true value $\beta$ in SM. This can be
seen explicitly from the following relation
 \beqa
\sin 2\beta_{J/\psi}
&=& { \sin 2\beta-\tilde{r}^B\sin 2 \phi^{B}
             \over \sqrt{ 1+2 \tilde{r}^B \cos 2(\phi^{B}+\beta)+ (\tilde{r}^B)^2}}.
\label{betajpsi}
\eeqa

The usual fit of the profile of the UT has also been made under
the assumption that there is no new physics beyond the SM.  The
angle ``$\beta_{fit}^{SM}$'' which is obtained from the
measurement of $K^0-\bar{K}^0$ and $B^0-\bar{B}^0$ mixing is
usually assumed to be the true angle $\beta$ associated with the
phase of $V^*_{td}$. In general, one should not make any
assumption for purpose of new physics probing, since it may be
greatly modified when new physics effects come into the evaluation
of $\epsilon_K$ and $\Delta m_{B_{(s)}}$.  As we have discussed
before even if we still use the same formulae as the ones in SM to
extract the angle $\beta$, we may only get the effective values of
$\rho_{fit}^{SM}$, $\eta_{fit}^{SM}$, and $\beta_{fit}^{SM}$.
Their values may be far away from the true ones, this is because
the experiment measured results for $\epsilon_K$ and $\Delta
m_{B_{(s)}}$ are the total ones which actually include all
possible new physics contributions if new physics truly exists.

 Combining  Eq.(\ref{newepsilon}) and  (\ref{newdmb}) and using
the definition of $\sin 2\beta$ in Eq.(\ref{angle-def}), we come
to the following simple relation
 \beqa
\sin 2\beta &=& C_K {1+r^B_{tt} \over 1+r^K_{tt}} \left[
\left|1-{\tilde{\epsilon}_K \over \epsilon_K }  \right|
      - \omega\bar{\eta} (1+r_{ct})   \right]
      \sqrt{ 1+2 \tilde{r}^B \cos 2(\phi^{B}+\beta)+ (\tilde{r}^B)^2},
\label{true-beta}
 \eeqa
 The relation between $\beta_{fit}^{SM}$ and $\beta$ is straight forward,
 \beqa
 \sin 2\beta_{fit}^{SM}=
     {  (1-\omega \bar{\eta}_{fit}^{SM})\,  \sin2\beta \over
      \sqrt{ 1+2 \tilde{r}^B \cos 2(\phi^{B}+\beta)+ (\tilde{r}^B)^2}\,
      [\left|1-\tilde{\epsilon}_K /\epsilon_K   \right|
      - \omega\bar{\eta} (1+r_{ct})] }
      \left( {1+r^K_{tt}\over 1+r^B_{tt} }\right)
\label{betafit}
 \eeqa

 To probe new physics effects, let us define the ratio $R_{\beta}$
between the two observables $\sin 2\beta _{J/\psi}$ and $\sin
2\beta_{fit}$  as follows
 \beqa
  R_{\beta}&=& {\sin
2\beta_{J/\psi}\over \sin 2\beta^{SM}_{fit}}  \nomb \\ &=&{1+r^B_{tt}
\over 1+r^K_{tt}}
              \left[{ \left|1-\tilde{\epsilon}_K /\epsilon_K   \right|
      - \omega\bar{\eta} (1+r_{ct}) \over 1-\omega \bar{\eta}^{SM}_{fit} }   \right]
    \left(1-\tilde{r}^B {\sin 2\phi_B \over \sin 2\beta}\right).
\eeqa
Note that the factor $ \sqrt{ 1+2 \tilde{r}^B \cos 2(\phi^{B}+\beta)+
(\tilde{r}^B)^2}$ in Eq.(\ref{true-beta}) and (\ref{betajpsi}) cancel
each other in the expression of $R_{\beta}$.  In the SM, $R_{\beta}$
is clearly equal to unity. From the above equation the deviation from
unity due to the new physics effects can be expressed as the products
of three terms:
(1) The ratio of new physics corrections to the $tt$ quark loop in the
box diagram between $B^0-\bar{B}^0$ and $K^0-\bar{K}^0$ mixings. Since
almost all the new physics models give similar contribution to
$B^0-\bar{B}^0$
and $K^0-\bar{K}^0$ mixing, their effect may cancel in some extent.
(2) The corrections to the $cc$ and $ct$-quark loops in the box
diagram in kaon meson as well as the ones from the additional
contributions to $K^0-\bar{K}^0$ mixing which contain new CP-violating
phases. The corrections from $cc$, $ct$ loops are suppressed due to the
small quark mass $m_c$ if the additional contributions come from the
new scalar particles such as the charged Higgs particle $H^\pm$ in some
models. However, in general case, the corrections may be considerable.
Although the small mass difference between $K^0$ and $\bar{K}^0$ has
imposed a strong constraint on the real part of $\tilde{M}_{12}$ , its
imaginary part can still be large and result in a sizable
$\tilde{\epsilon}_K$ such as the superweak-type interactions.
(3) The corrections from the additional contributions to
$B^0-\bar{B}^0$ with new CP-violating phases. It can be seen
that the corrections depend on the product of $\tilde{r}^B$ and $ \sin
\phi_B$. If $\phi^B=0$ the correction to $R_{\beta}$ disappears.
 This is directly related to the cancelation of the factor
$ \sqrt{ 1+2 \tilde{r}^B \cos 2(\phi^{B}+\beta)+ (\tilde{r}^B)^2}$ in
the ratio $R_{\beta}$. Although the new physics with zero $\phi_B$ can affect
both $\beta_{J/\psi}$ and $\beta_{fit}$, their net effect on $R_{\beta}$
is diminished.
%

\section{$R_{\beta}$ and  $\beta^{SM}_{fit}$ from the global  fit}

To get the value of $R_{\beta}$ from the data, one needs to know the
values of $\beta_{J/\psi}$ and $\beta^{SM}_{fit}$.  Unlike the measurement
of $\beta_{J/\psi}$ which is theoretically clean, the extraction of
$\beta^{SM}_{fit}$ seems to surfer from some uncertainties as it involves parameters with
large theoretical uncertainties.  There is a long time debate on the
treatment of the theoretical errors.  Basicly there are three different
approaches\cite{ciuch}.

(1) Assuming all the errors, both experimental and theoretical ones are the
random variables which obey Gaussian distribution \cite{london}.  The
most recent fitted results at $95\%CL$ are \cite{x.g.he}:
\beqa
-0.82\leq \sin 2\alpha^{SM}_{fit} \leq0.42, \;\;
0.49\leq\sin 2\beta^{SM}_{fit} \leq0.94, \;\;
42^\circ\leq \gamma^{SM}_{fit} \leq 83^\circ
\eeqa

(2) The distribution of experimental error is treated as Gaussian,
but one imposes a $prior$ flat distribution on the theoretical error,
i.e. assuming the theoretical parameters are the random variables which
are uniformly populated in some reasonable regions , then
by using the Bayesian approach, the distributions of the parameters can be
determined \cite{ciuch,flat}.
The recent fits using this
method give the results \cite{ciuch}
\beqa
-0.88\leq \sin 2\alpha^{SM}_{fit} \leq0.04, \;\;
0.57\leq\sin 2\beta^{SM}_{fit} \leq0.83, \;\;
42^\circ\leq \gamma^{SM}_{fit} \leq 67^\circ
\eeqa

(3) The distribution of experimental error is treated as Gaussian,
but one $does \ not $ assume any distribution on the theoretical error,
since they are unknown parameters which should have a unique
value. their errors should be largely deduced with the improvement of
the theory and the experiments.  To get a quantitative result, the
space of the allowed region for the theoretical parameters are scanned.
For each set of parameters a contour at some confidence level (
for example $68\%$ or $ 95\%$) is made, then the whole region
enveloping all the contours is considered as the allowed region at
some $overall $ confidence level.  Note that in doing this ,  the
final results will have no clear probability explanation.  The most
recent fit gives \cite{scanning1,scanning2}
\beqa
-0.95\leq \sin 2\alpha^{SM}_{fit} \leq 0.5,\;\;
0.5\leq\sin 2\beta^{SM}_{fit} \leq0.85, \;\;
40^\circ \leq \gamma^{SM}_{fit} \leq 84^\circ
\eeqa

In this paper, to get a more conservative conclusion, we adopt the last
method and update the fit with the latest data on $\Delta m_B$ and
$\Delta m_{B_s}$

Let us briefly describe the method  and parameters  used in the
fitting of the unitarity triangle (UT). The  detailed description of the fitting procedure
can  be found in Ref.\cite{babarbook}.

Among the four Wolfenstein parameters $A, \rho, \eta$ and $\lambda$,
the value of $\lambda$ has been well determined with a relative high precision
through semileptonic kaon decays, $K^+\to \pi^0e^+\nu_e$ and
$K^0_L \to \pi^-e^+\nu_e$.   In the fit, we quote the value of
$\lambda=0.2196$\cite{PDG}  and take it as a fixed parameter.

The constraints on the apex of UT can be obtained from various
experiments of the semileptonic $B$ decays. In $b\to c$ transition
such as $B\to X_c l \nu$ and $\overline{B^0}\to D^{*+} l^+
\bar{\nu _e}$, the value of $|V_{cb}|$ can be determined. Here we
quote the value of $V_{cb}$ from the most recent LEP
results\cite{LEPvcb}
\beqa
 |V_{cb}|=(40.4\pm 1.8)\times 10^{-3}, \quad (A=0.850\pm 0.037)
\eeqa
which is an average of measurements between inclusive and
exclusive $B$ decays. From $B\to \rho l^+\bar{\nu_e}$ the value of
$|V_{ub}|$ could be extracted.  However, in the determination of
$V_{ub}$, some model dependences have to be involved in the
evaluation of the ratio $|V_{ub}/V_{cb}|$, which results in a
considerable theoretical error. For a recent review of the
theoretical error in determining $|V_{ub}/V_{cb}|$ , we refer to
the reference \cite{ural}.  The recent measurements of $V_{ub}$
from LEP and CLEO collaboration give the following
values:\cite{LEPvub,CLEOvub}
 \beqa
|V_{ub}|&=&(4.13^{+0.42}_{-0.47}(\mbox{stat.+det.})^{0.43}_{-0.48}
               (b\to c \mbox{ syst.})^{+0.24}_{-0.25}(b\to u \mbox{ sys.})\nomb\\
              &&\pm 0.02 (\tau_b)\pm 0.20  (\mbox{Model})  \times 10^{-3}. (\mbox{LEP})\\
|V_{ub}|&=&(3.25\pm 0.14(\mbox{stat.})^{+0.21}_{-0.29}
               (\mbox{syst.})\pm0.55 (\mbox{model})  \times 10^{-3}. (\mbox{CLEO})
\eeqa
 Note that both the determination of $|V_{ub}|$ and
$|V_{cb}|$ may still suffer from sizable theoretical
uncertainties\cite{WWY1,WWY2,WY}. In the fit, we take the central
value of $|V_{ub}/V_{cb}|$ as a free theoretical parameter which
lies in the range
 \beqa
 <V_{ub}/V_{cb}>\in [0.07,0.1]
  \eeqa
 and the  value of $|V_{ub}|$ has the following form
 \beqa
|V_{ub}|=<V_{ub}/V_{cb}> |V_{cb}|\pm 0.14
 \eeqa
It is noted that possible new physics contributions to the
semileptonic bottom meson decays are in general expected to be
small,  the extracted CKM parameters $|V_{cb}|$ and $|V_{ub}|$ are
regarded as the true ones $|V_{cb}|=A\lambda^2$ and
$|V_{ub}|=\sqrt{\rho^2 + \eta^2}$.

The other important constraint comes from the measurements of
$\epsilon_K$. The expression of $\epsilon_K$ has been discussed in
the previous section. From Eq.(\ref{newepsilon}), the major
theoretical error comes from the hadronic parameter $\hat{B}_K$,
and the QCD correction $\eta_1$, and  $\eta_3$. the recent lattice
calculations give the result \cite{BK}
 \beqa
0.8 \leq \hat{B}_K  \leq 1.1 .
 \eeqa
 The errors of $\eta_i$ can be
found in the previous section. There are also experimental errors
on the values of $c$ and $t$ running quark masses,
\beqa
\bar{m}_c(m_c)=1.25\pm 0.1 \;\mbox{GeV}, \;\;\;
\bar{m}_t(m_t)=165.0\pm 5.0 \;\mbox { GeV}
\eeqa

The constraint of $R_t$ could arises from both $B^0-\bar{B}^0$ and
$B^0_{s}-\bar{B}^0_{s}$ mixing. Here the parameter with a large
uncertainty is $f_B\sqrt{B_B}$. In the fit, we take the
value\cite{fB}
 \beqa
160 \;\mbox{MeV} \leq f_B\sqrt{B_B} \leq
240\; \mbox{MeV}
 \eeqa

Recently, the BaBaR and Belle Collaboration has reported their first measurements on
$\Delta m_{B_d}$. The combined result from the hadronic and
semileptonic decays is \cite{babar,belledmb} :
\beqa
\Delta m_{B_d}&=&0.512 \pm 0.017(\mbox{stat.}) \pm0.022 (\mbox{syst.}) \;ps^{-1} \;\; \mbox{(BaBaR)} \nomb\\
\Delta m_{B_d}&=&0.456 \pm 0.008(\mbox{stat.}) \pm0.030 (\mbox{syst.}) \;ps^{-1} \;\; \mbox{(Belle)}
\eeqa
comparing with the old world average value of $0.472\pm0.017 ps^{-1}$ \cite{PDG},
the result of BaBaR is slightly  higher. Now the new world average is
\beqa
\Delta m_{B_d}=0.478\pm0.013 \; ps^{-1}
\eeqa

Besides $\Delta m_B$,  the mass difference of $B^0_{s}$ and $\bar{B}^0_{s}$
also imposes a strong constraint on the size of $R_t$. It is helpful to introduce
a $SU(3)$ breaking factor
$\xi^2\equiv(f_{B_s}\sqrt{B_{B_s}})^2/(f_{B}\sqrt{B_{B}})^2$ ,
the constraint on $R_t$ from $\Delta m_{B_s}$ reads
\beqa
R_t \leq {\xi \over \lambda}\sqrt{
\frac{m_{B_s}}{m_{B_d}}
\frac{\Delta m_{B_d}}{\Delta m_{B_s}|_{min}} }
\eeqa
In the fit , we take $\xi \in [1.08,1.2]$ \cite{BK}.
The lower bound of $\Delta m_{B_s}$ is now updated to $\Delta m_{B_s}>
14.9\ \mbox{ps}^{-1}$ at $95\%$CL \cite{LEPsite} which is also higher
than the previous value of $14.3$ ps$^{-1}$.  This bound is obtained
through the so called 'amplitude method' \cite{moser}. The latest data
 indicate that $\Delta m_{B_s}\sim 17.7 $ ps$^{-1}$. If
this result can be confirmed by the future experiment, $\Delta
m_{B_s}$ will impose strong constraint on  new physics models
\cite{london-dms}.

The other two important CP-violation observables are the direct CP
violation parameter $\epsilon'/\epsilon$ and $\sin 2\beta$ from
the time dependent CP asymmetry of decay $B\to J/\psi K_S$.  As
the measurements of $\sin 2\beta$ and $\epsilon'/\epsilon$ are now
preliminary, they are not considered in our present fitting .
However, it needs to emphasize that including the constraints from
$\epsilon'/\epsilon$ may impose a nontrivial lower bound of $\eta
>0.32$ \cite{scanning1}. It would be interesting to further
discuss its constraint with the improved prediction for
$\varepsilon'/\varepsilon$ \cite{wu-eps}.

The basic idea of the fit is based on the least square method.
 The first step of the procedure is to construct a quantity $\chi^2$ which
has the following form
 \beqa
\chi^2=\sum_{i}\frac{(f_i(A,\rho,\eta)-<f_i>)^2}{\sigma_i^2}
\eeqa
where $<f_i>$ and $\sigma_i$ are the central values and
corresponding errors of the experimentally measured observables.
They contain $|V_{cb}|, |V_{ub}|, |\epsilon_K|,\Delta m_{B}$ and
$m_c, m_t$. $f_i(A,\rho, \eta)$ are the values calculated from the
theories. They are the functions of parameters $A,\rho, \eta$. The
set of $A, \rho, \eta$ which minimizes the $\chi^2$ will be
regarded as the best estimated values. On the determination of
$\Delta m_{B_s}$, the ' amplitude method' \cite{moser} is adopted,
 which uses the amplitude curves measured from the experiments
\cite{LEPsite} and add a term $\chi^2_s$ to the total $\chi^2$:
\beqa
 \chi^2_s=\frac{(1-{\cal A}(\Delta m_{B_s}))^2}{\sigma^2_{\cal A}
              (\Delta m_{B_s})}
\label{amplitude}
\eeqa

  Note that  there is  an alternative approach to build the log-likelihood
function in which one uses the values relative to the ones at
$\Delta m_s=\infty$, i.e. use $[\frac{\left( {\cal A}-1
\right)^2}{\sigma_A^2}
     -\frac{ {\cal{A}}^2}{\sigma_A^2}]$ instead of
$\frac{\left( {\cal A}-1 \right)^2}{\sigma_A^2}$
 \cite{moser,ciuch,piotto}. The detailed
comparison between these two approaches can be found in
Ref.\cite{ciuch}. In the fit, we choose $A, \rho, \eta $ and $m_c,
m_t$ as free parameters to be fitted  and scan the allowed range
for the theoretical parameters $<V_{ub}/V_{cb}>$, $\hat{B}_K,
f_B\sqrt{B_B}, \eta_s$, $\eta_1$, and $\eta_3$. For each set of
theoretical parameters, a $68\%(95\%)$ CL contour which
corresponds to $\chi^2=\chi^2_{min}+ 2.4(6.0)$ is made, then the
region enveloping all the contours is the allowed region for
$\rho$ and $\eta$ at an $overall \ 68\%(95\%)$ level.  As usual, a
$\chi^2$ probability cut is used to reject the contours with
relative high $\chi^2_{min}$ which means that the fit is not
consistent.

\begin{figure}
\centerline{ \psfig{figure=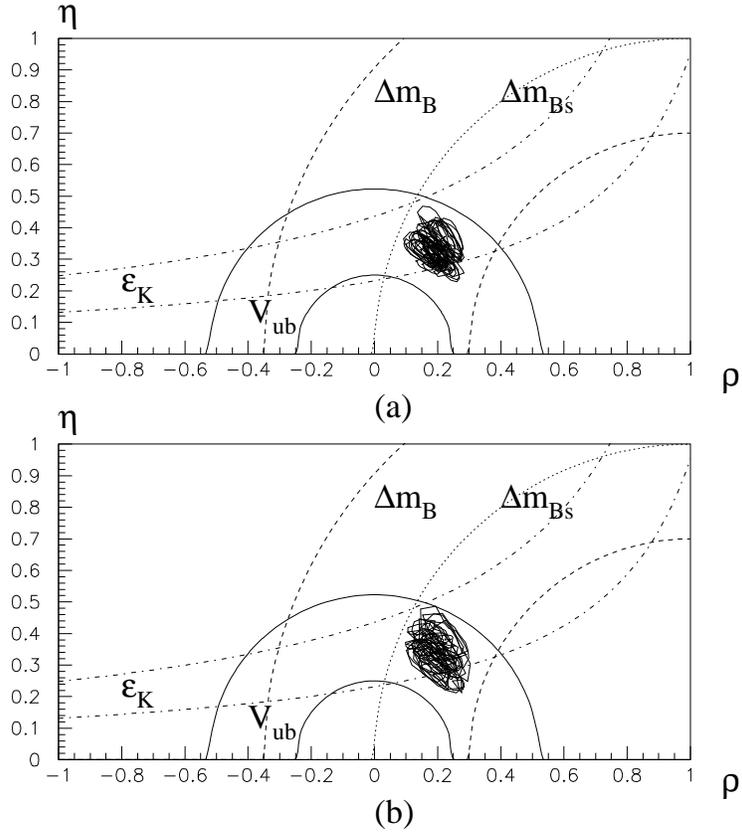, width=12cm} }
 \caption{The allowed region for $\rho$ and $\eta$ from the global fit at 68$\%$CL (a)
and 95$\%$CL (b).  The individual constraints from $V_{ub}$, $\epsilon_K$ and
$B^0_{(s)}-\bar{B}^0_{(s)}$ are also shown.   }
\label{contour.eps}
 \end{figure}

The fit is implemented by using the program package
MINUIT\cite{miniut}. The result is shown in Fig.\ref{contour.eps}.
 From the figure, one can read off the allowed region for
$\rho$ and $\eta$.  The results are
 \beqa
& &  0.09\leq \rho^{SM}_{fit} \leq 0.28\, , \quad 0.23 \leq \eta^{SM}_{fit}
\leq 0.46 \quad (at\ 68\%\ CL) \nonumber \\
  & &  0.10\leq \rho^{SM}_{fit} \leq 0.30\, , \quad 0.20 \leq \eta^{SM}_{fit} \leq 0.49 \quad (at\
95\%\ CL)
 \eeqa

or equivalently,
  \beqa
& & -0.89 \leq \sin 2\alpha^{SM}_{fit}  \leq 0.37, \ \ 0.56 \leq \sin
2\beta^{SM}_{fit}  \leq 0.85,  \ \ 41^\circ \leq \sin^2 \gamma^{SM}_{fit}
\leq 75^\circ ,  \quad (at\ 68\%\ CL) \nonumber \\
  & & -0.95 \leq \sin 2\alpha^{SM}_{fit}
\leq 0.41, \ \ 0.48 \leq \sin 2\beta^{SM}_{fit}  \leq 0.88,  \ \
 38^\circ \leq \sin^2 \gamma^{SM}_{fit} \leq 76^\circ , \quad
 (at\ 95\%\ CL)
 \eeqa
and the range for the combination factor of the CKM matrix
elements, $Im \lambda_t A^2\lambda^5\ \eta$, used in the
calculation of direct CP-violating parameter
$\varepsilon'/\varepsilon$ in kaon decay\cite{wu-eps}, is given by
 \beqa
 & & 0.76\times 10^{-4} \leq Im \lambda_t = A^2\lambda^5\ \eta \leq 1.73\times
 10^{-4}\ , \quad (at\ 68\%\ CL)  \nonumber \\
 & & 0.66\times 10^{-4} \leq Im \lambda_t = A^2\lambda^5\ \eta \leq 1.84\times
 10^{-4}\ , \quad (at\ 95\%\ CL)
 \eeqa

  As the allowed ranges for theoretical errors which are the
main sources of total errors are the same in both fits ,
the results at 68$\%$ and 95$\%$CL are quite similar,
It is needed to note that by using the scanning method, there is
no  probability explanation of the contours. One can not get the
usual central value of the fitted parameters.

In  the fitting, the $\chi^2$ probability cut is set to be
$Prob(\chi^2) \leq 5\%$.  The final results depend on the cut. We
have checked that a lower cut such as $2\%$ or $1\%$ will give a
larger allowed range, which allows the angle $\gamma$ to be
greater than $90^\circ$.  The probability of large $\gamma$ has
been widely discussed in Refs..\cite{large-gamma} to meet the
recent CLEO data on hadronic charmless $B$ decays. However, the
model independent analysis show that in general there are two
solutions for $\gamma$ from the charmless $B$ decays. The one with
$\gamma<90^\circ$ and the other one with $\gamma>90^\circ$
\cite{wu-zhou-Bdecay}. The data of $\Delta m_B$ especially on
$\Delta m_{B_s}$ may strongly constrain $\gamma$ to be less than
$90^\circ$. From Fig.\ref{contour.eps}, with the cut of $5\%$,
there is no indication of $\gamma>90^\circ$.

With the angle $\beta_{fit}$ obtained from the above fit and the
 average value for $\beta_{J/\psi}$, the ratio $R_{\beta}$
 is found to be
  \beqa
   0.57 \leq R_{\beta}^{exp.} \leq 1.1\, \quad (at\ 68\%\ CL)\ , \quad
   0.31 \leq R_{\beta}^{exp.} \leq 1.36\, \quad (at\ 95\%\ CL)
\label{Rfit}
  \eeqa
The present data prefer a small value of $R_{\beta}<1$ at
1-$\sigma$ level. If it is confirmed by the future experiments, it
will of course be a signal of new physics. Further more,
$R_{\beta}$ may be used to distinguish different effects of new
physics. This will be discussed in the next section.

 \section{ new physics effects on the ratio $R_{\beta}$ }

The contributions of new physics to $R_{\beta}$ depend on the models. In this
section, we choose two type of new physics models  to
illustrate  new physics effects.  one of them is the model with MFV,
the other is the models with new CP-violating  phases, such as the
S2HDM.
\subsection{$R_{\beta}$ in models with MFV}

In a class of the models with MFV, the expression of $R_{\beta}$
may be greatly simplified.  As MFV implies that there are no new
operators beyond the SM, the values of $r^B_{tt}$ and $r^K_{tt}$
which come from the same internal $tt$ quark loop in
$B^0-\bar{B}^0$ and $K^0-\bar{K}^0$ mixings must be the same
\cite{buras1,buras2,buras-mfv}. MFV also means that new
CP-violating phases must vanish. Thus the following conditions
hold \beqa r^B_{tt}=r^K_{tt},  \;\; \tilde{r}^B=0,\;\;
\phi^B=0,\;\; \mbox{and} \;\; \tilde{\epsilon}_K=0 \eeqa Thus one
has

\beqa
\sin 2\beta_{J/\psi}=\sin 2\beta,
\label{beta in MFV}
\eeqa
which shows that in this case the angle $\beta_{J/\psi}$
measured from $B\to J/\psi K_S$ is the 'true' angle $\beta$ up to a
four fold ambiguity.  Assuming that $r_{ct}$ is small, which is a good
approximation for many models \cite{buras-mfv}, the ratio  $R_{\beta}$ has the following simple
form
\beqa
\left. R_{\beta}\right|_{MFV}={1-\omega \bar{\eta}\over
 1-\omega \bar{\eta}^{SM}_{fit}}
\eeqa

Since the value of $|V_{ub}|$ is extracted from the semileptonic
$B\to \pi(\rho) e \nu_{e}$
decay  dominated by the tree diagrams, it is widely believed
that its value is not likely to be affected by new physics. This may
result in the relation:
$\bar{\rho}^2+\bar{\eta}^2=(\bar{\rho}^{SM}_{fit})^2+(\bar{\eta}^{SM}_{fit})^2=R_u^2$.
Combining this relation with Eqs.(\ref{beta in MFV}) and
(\ref{angle-def}), the values of $\bar{\rho}$ and $\bar{\eta}$ can be solved as
functions of two obserbales $R_u$ and $\beta_{J/\psi}$.
\beqa
\bar{\rho}=R_u \cos \varphi_{\pm}, \;\;\;
\bar{\eta}=R_u \sin \varphi_{\pm} .
\eeqa
with $\varphi_{\pm}= \sin^{-1} ( {\sin \beta_{J/\psi} / R_u} )
                           \pm \beta_{J/\psi}$.  Note that the
solution for $\varphi_{+}$ corresponds to $\bar{\rho}< 0$. This is
quite different from the value of $\bar{\rho}^{SM}_{fit}$ which
seems be positive under the constraint from the data of
$B^0_{s}-\bar{B}^0_{s}$ mixing. If one assumes that the difference
between true $\bar{\rho}$ and the fitted $\bar{\rho}^{SM}_{fit}$
caused by new physics effects is not too large, the solution for
$\varphi_{+}$ could be ignored. Taking the value of $R_u$ and
$\sin 2\beta_{J/\psi}$, one may find at 1$\sigma$ level

 \beqa
  1.03\leq R_{\beta}|_{MFV}\leq 1.36\,  \mbox{ for } \varphi_-\, .
 \eeqa
 For $\varphi_+$ solution, one gets a small value for $ R_{\beta}|_{MFV}$, i.e.,
 $ \,0.90\leq R_{\beta}|_{MFV}\leq 1.21$. In obtaining these numerical values,
the errors of $\omega$ and $\bar{\eta}^{SM}_{fit}$ are considered
to be independent for simplicity. In general, they may be treated
to be correlated as they  all depend on the hadronic parameters.
For a more detailed analysis, it should be interesting to include
the correlation effects.

Comparing $R_{\beta}|_{MFV}$ with the experimental values of $R^{exp.}_{\beta}$ in
Eq.(\ref{Rfit}) , it is found that the models with MFV prefer to
give a large value of $R_{\beta}$ which is not likely to  agree
with  the current data. If the disagreement is confirmed by the
future more precise data, all the models with MFV will be ruled
out. It implies that new physics effects with flavor changing
interactions beyond the CKM quark mixing must be involved to
explain the data.

\subsection{$R_{\beta}$ in S2HDM}
\begin{figure}
\centerline{ \psfig{figure=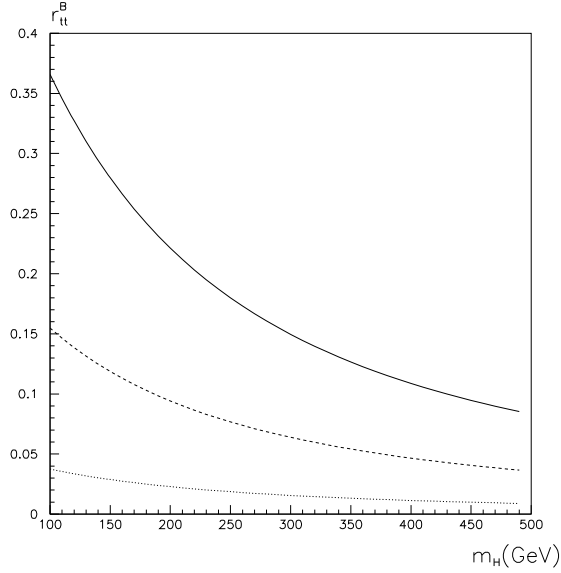, width=8cm} }
 \caption{$r^B_{tt}$ as a function of the charge Higgs mass $m_{H^\pm}$.
The three cures correspond to $\xi_t$=0.6(solid),0.4(dash),0.2(dotted)
respectively.}
\label{r1.eps}
 \end{figure}

The S2HDM without imposing discrete
symmetries\cite{TH1,TH2,TH3,TH4,WW,YLW,Wu-Zhou,soni-etc} is a good
example for models which can give nonzero value of
$\tilde{\epsilon}_K,\tilde{r}^B$ and $\phi^B$. In the S2HDM, the
tree level flavor changing neutral current can be induced from the
scalar interactions between neutral Higgs bosons and quarks. As
all the Yukawa couplings in the model are allowed to be complex,
the S2HDM  can give rich sources of CP
violation\cite{WW,YLW,keung}. It is not difficult to find that the
values of $r^B_{tt}$, $ \tilde{r}^B$ and $\phi^B$ in S2HDM are
given by \beqa r^B_{tt}&=&
    \frac{1}{4}\abs{\xi_t}^4 y_t\frac{\eta^{HH}_{tt}}{\eta_{B}}
      \frac{ B^{HH}_V(y_t)}{B^{WW}(x_t)}
   +2 \abs{\xi_t}^2 y_t \frac{\eta^{HW}_{tt}}{\eta_{B}}
      \frac{ B^{HW}_V(y_t,y_W)}{B^{WW}(x_t)} \\
\tilde{r}^B \sin 2\phi^B&=&\frac{\tilde{B_B}}{B_B}\sum_{k}
    \lb\frac{2\sqrt{3}\pi v m_B}{m_{H^0_{k}} m_t} \rb^2 \frac{m_d}{m_b}
    \frac{1}{V^2_{td}} \frac{\mbox{Im}(Y^d_{k,13})^2}{\eta_{B}B^{WW}(x_t)}
\eeqa
where $\xi_q$ is the Yukawa coupling constant between charged Higgs
and $q$ quark.  $B_{(V)}^{WW(HW,HH)}$ are the integral functions of the
box diagrams \cite{wise} with variable $x_t=m_t^2/m^2_W$, $y_t=m_t^2/m^2_H$
 and $y_W=m_W^2/m^2_H$. $\eta^{HH}_{tt}$ is the QCD correction. It is
seen that  the value of $r^B_{tt}$ is directly related to the coupling
$\xi_t$. In Fig.\ref{r1.eps}, $r^B_{tt}$ is plotted as a function of the
charged Higgs mass $m_{H^{\pm}}$ with different values of $\xi_t$. It
can be seen that the typical value of $r^B_{tt}$ is around $0.2$. For
large $\xi_t$ , $r^B_{tt}$ could become large.

The value of $\tilde{r}^B$ depends on the imaginary part of $Y^f_{k,ij}$ which
is given by reduced Yukawa couplings between the $k$-th neutral Higgs boson and
quarks. $\tilde{B}_B$ and $B_B$ are the bag parameters for
$(S+P)\bigotimes(S-P)$ and $(V-A)\bigotimes(V-A)$ four quark operators
respectively.

Similar calculations can be made for the $K^0-\bar{K}^0$ mixing.
$r^K_{tt}, r_{ct}$   in S2HDM are given by
 \beqa
r^K_{tt}&=& r^B_{tt} \\
r_{ct}&=&
{ \sqrt{y_t y_c} ({x_t \over x_c}) {1 \over2} \eta^{HH}_{tt} |\xi_c\xi_t|^2
  B^{HH}_{V}(y_c,y_t) +4 \eta^{HW}_{ct} Re(\xi_c\xi_t) B^{HW}_V(y_c,y_t,y_W)
                                    \over
 ({x_t \over x_c})\eta_B B^{WW}(x_c,x_t) -\eta_1 B^{WW}(x_c) } \\
 \eeqa

Since $x_t/x_c$ is of order ${\cal O}(10^4)$, $r^K_{ct}$ can be approximately
written as
\beqa
r^K_{ct}\approx {1 \over 2} \sqrt{y_t y_c} {\eta^{HH}_{tt} \over \eta_B}
{  B^{HH}_{V}(y_c,y_t) \over B^{WW}(x_c,x_t)  } |\xi_c\xi_t|^2
\eeqa
As the value of $\sqrt{y_t y_c}$ is of  order $10^{-2}$ and $
|\xi_c\xi_t|^2 $ is typicly of order $1$, one may find
that $r_{ct}$ is relatively small and of  order ${\cal
O}(10^{-2}\sim 10^{-1}) $.
   The S2HDM contribution to $\tilde{\epsilon}_K$ is complicate,
it may arise from both short and long distance interactions. The
detailed discussion of $\tilde{\epsilon}_K$ can be found in Ref.
\cite{Wu-Zhou}

\begin{figure}
\centerline{ \psfig{figure=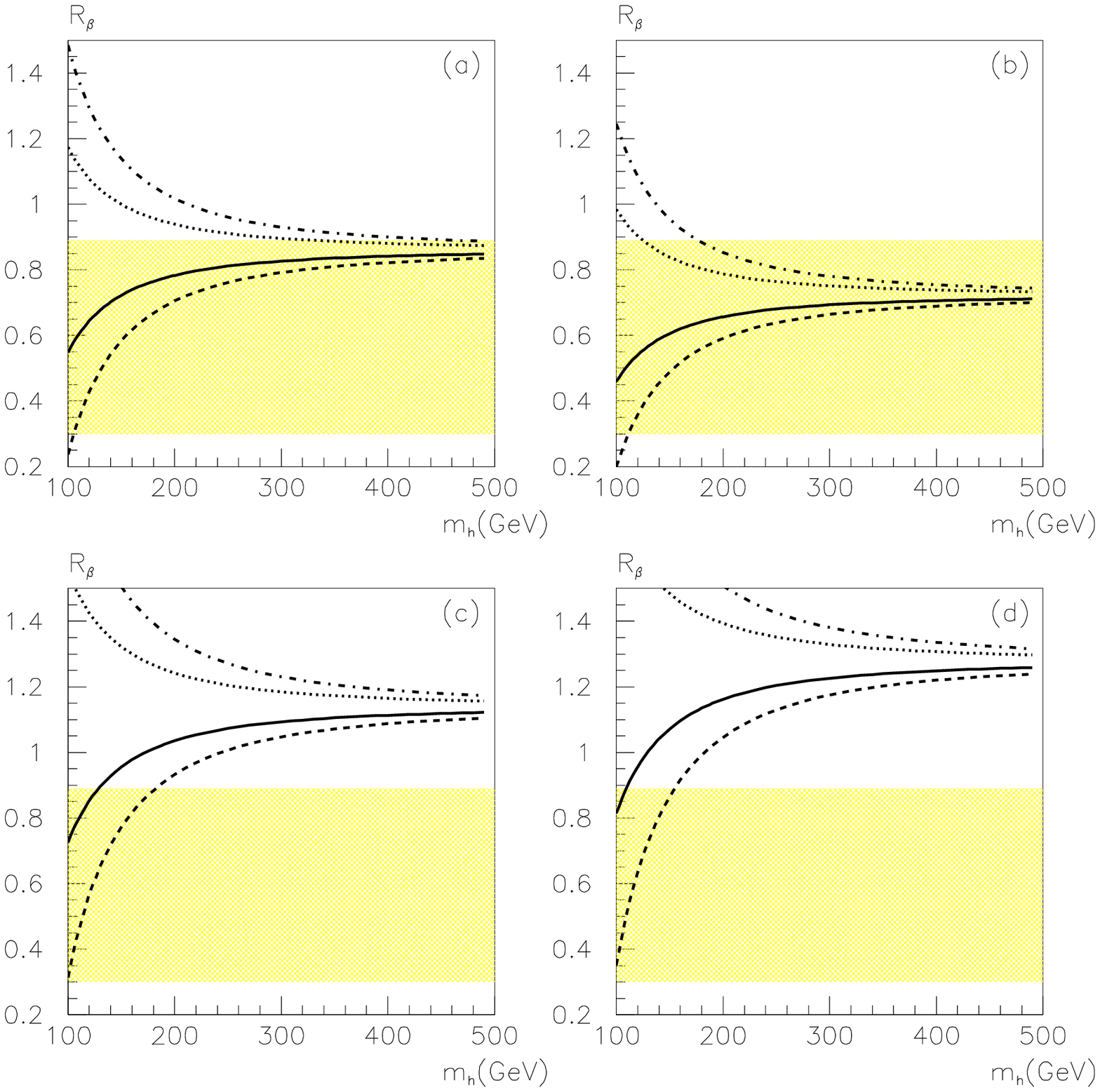, width=8cm} }
 \caption{$R_{\beta}$ as a function
of the mass of neutral Higgs mass $m_{h}$ with different values of
 $Im(Y^d_{1,13})^2$ and $\tilde{\epsilon_K}/\epsilon_K$. The shadowed
 band indicates the allowed region for $R_{\beta}$ at 68$\%$ CL.\\
(a) $Im(Y^d_{1,13})^2$=0.01(solid), 0.02 (dashed), $-$0.01
(dotted), $-$0.02 (dotted-dashed) with
$|1-\tilde{\epsilon_K}/\epsilon_K|=0.9$.\\ (b) The same as (a)
with $|1-\tilde{\epsilon_K}/\epsilon_K|=0.8$.\\ (c) The same as
(a) with $|1-\tilde{\epsilon_K}/\epsilon_K|=1.1$.\\ (d) The same
as (a) with $|1-\tilde{\epsilon_K}/\epsilon_K|=1.2$. }
\label{r.eps}
 \end{figure}

If one ignores the $r^K_{ct}$ , the ratio
$R_{\beta}$ mainly depends on the value of $\tilde{\epsilon}_K$
, $Im Y^d_{k,13}$ and $ m_{H^0}$
as well as the true value of $\sin 2 \beta$ and $\bar{\eta}$.  The ratio
$R_{\beta}$ has the following form

\beqa
R_{\beta}={|1-\tilde{\epsilon_K}/\epsilon_K|-\omega \bar{\eta}\over
 1-\omega \bar{\eta}^{SM}_{fit}}
\left(1-
\frac{\tilde{B_B}}{B_B}\sum_{k}
    \lb\frac{2\sqrt{3}\pi v m_B}{m_{H^0_{k}} m_t} \rb^2 \frac{m_d}{m_b}
    \frac{1}{|V^2_{td}|} \frac{\mbox{Im}(Y^d_{k,13})^2}{\eta_{B}B^{WW}(x_t)}
\right)
 \eeqa
Which shows that the new interactions between neutral Higgs and
quarks are the main sources for the changing of ratio $R_{\beta}$.
The value of $R_{\beta}$ as the function of neutral Higgs mass
$m_{H^0}$ is plotted in Fig.\ref{r.eps} with different values of
$Im Y^d_{k,13}$ and $\tilde{\epsilon}_K/\epsilon_K$. Where for
simplicity we have taken $\bar{\eta} \simeq \bar{\eta}_{fit}=0.35$
and $\tilde{B}_B \simeq B_B$.  From the figure, the value of
$R_{\beta}$ can be smaller than 1 for positive $Im(Y^d_{K,13})^2$
or $\tilde{\epsilon}_K/\epsilon_K$. Thus the present data can be
easily explained in the S2HDM.

\section{ conclusions }
In conclusion, we have investigated the measurement of the UT
angle $\beta$ within and beyond the SM. It has been shown that if
new physics beyond the SM truly exists, the angle $\beta$
extracted from the global fit(denoted by $\beta^{SM}_{fit}$) and
the one extracted from $B\to J/\psi K_S$ (denoted by
$\beta_{J/\psi}$) could be in general all different from the
'true' angle $\beta$ in the SM. The observable $R_{\beta}$, which
is the ratio between the angle $\beta$ measured from the time
dependent CP asymmetry of decay $B\to J/\psi K_S$ and the one
extracted from the global fit, is introduced and studied in
detail. By using the scanning method, the value of $R_{\beta}$ is
obtained from an update of the fit of the UT with the latest data
and it is found to lie in the range.
 \[
   0.57 \leq R_{\beta}^{exp.} \leq 1.1\, \quad (at\ 68\%\ CL)\ , \quad
   0.31 \leq R_{\beta}^{exp.} \leq 1.36\, \quad (at\ 95\%\ CL)
  \]
 As the SM calculation gives $R_{\beta}=1$, the deviation of
 $R_{\beta}$ from unity is a clean signal of new physics.  On the
study of new physics effects, we have given a general
parameterization of $R_{\beta}$. By using this parameterization,
the effects from the models with MFV and the S2HDM have been
investigated in detail. It has been found that the MFV models seem
to give  a large $R_{\beta}$ relative to the current data.
If it can be comfirmed by the future experiments,
it will  indicate that this kind of new physics are difficult to explain the
small value of $R_{\beta}$. Such models
include type I and type II 2HDM and some simple versions of
SUSY. Thus new physics with additional phases should be considered.
The  most  recent  discussions on models with MFV.
can also be found in Ref.\cite{Ali,buras3} .
As an illustration for those models with additional phases beyond
the one in the CKM matrix element, the S2HDM effects on
$R_{\beta}$ has been studied, especially, if the superweak type
contributions\cite{LW1} to $\epsilon_K$ are considered, it could
easily coincide with the data.

\centerline{ \bf acknowledgments } One of the authors (Y.L.Wu)
would like to thank L. Wolfenstein for helpful discussions.  This
work is supported in part by NSFC under grant No. 19625514 and
Chinese Academy of Sciences.


\begin{figure}
\centerline{ \psfig{figure=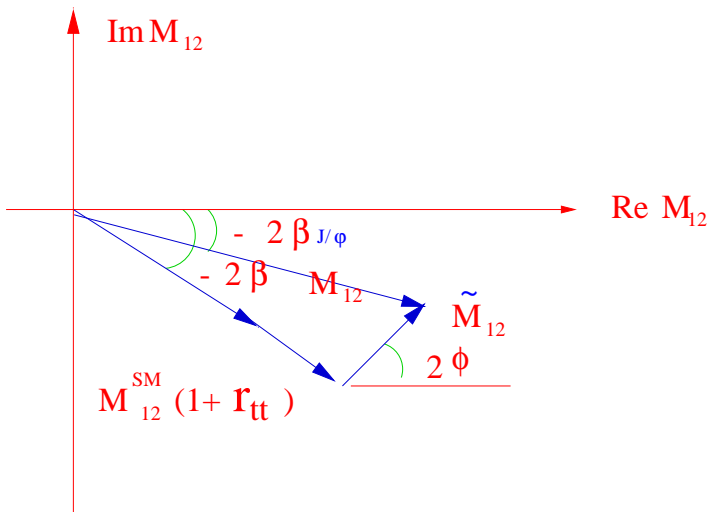, width=12cm} }
 \caption{The new physics contribution to $M_{12}$}
\label{newphys.eps}

 \end{figure}

\newpage

\newpage

\end{document}